\documentclass[journal]{IEEEtran}
 
\ifCLASSINFOpdf
\else
\fi

\usepackage{graphicx}
\usepackage{tabularx}
\usepackage{algorithm} 
\usepackage{algorithmic} 
\usepackage{multirow} 
\usepackage{amsmath} 
\usepackage{xcolor}
\usepackage{mathrsfs}
\usepackage{amsmath}
\usepackage{bbding}
\usepackage{pifont}
\usepackage{wasysym}
\usepackage{amssymb}
\usepackage{subfigure}
\usepackage{booktabs}
\usepackage{cite}
\usepackage{float}
\usepackage{indentfirst}
\usepackage{float}
\usepackage{changepage} 
\usepackage{booktabs}

\begin{document}


\title{Multi-Sample based Contrastive Loss for Top-k Recommendation}

\author{ 
Hao~Tang, 
Guoshuai~Zhao, 
Yuxia~Wu,
and Xueming~Qian

\thanks{ 
H. Tang, Y. Wu is with the School of Information and Communication Engineering, Xi’an Jiaotong University, Xi’an 710049, China (e-mail: th1002@stu.xjtu.edu.cn, wuyuxia@stu.xjtu.edu.cn ). 

G. Zhao (corresponding author) is with the School of Software Engineering, Xi'an Jiaotong University, Xi'an 710049, China (e-mail: guoshuai.zhao@xjtu.edu.cn ). 

X. Qian  is with the Ministry of Education Key Laboratory for Intelligent Networks and Network Security, School of Information and Communication Engineering, and SMILES LAB, Xi’an Jiaotong University, Xi’an 710049, China. (e-mail: qianxm@mail.xjtu.edu.cn). }

\thanks{ }}
 
\markboth{Journal of \LaTeX\ Class Files, ~Vol. ~14, No. ~8, August~2015}%
{Shell \MakeLowercase{\textit{et al. }}: Bare Demo of IEEEtran. cls for IEEE Journals}

\maketitle

\begin{abstract}

The top-k recommendation is a fundamental task in recommendation systems which is generally learned by comparing positive and negative pairs.  
The Contrastive Loss (CL)  is the key in contrastive learning that has received more attention recently and we find it is well suited for top-k recommendations. However, it is a problem that CL treats the importance of the positive and negative samples as the same. On the one hand, CL faces the imbalance problem of one positive sample and many negative samples. On the other hand, positive items are  so few  in  sparser datasets that their importance should be emphasized.  Moreover, the other important issue is that the sparse positive items are still not sufficiently utilized in recommendations.  So we propose a new data augmentation method by using multiple positive items (or samples) simultaneously with the CL loss function. 
Therefore, we propose a Multi-Sample based Contrastive Loss (MSCL) function which solves the two problems by balancing the importance of positive and negative samples and data augmentation. And based on the graph convolution network (GCN) method, experimental results demonstrate the state-of-the-art performance of MSCL. The proposed MSCL is simple and can be applied in many  methods. We will release our code on GitHub upon the acceptance.  

\end{abstract}

\begin{IEEEkeywords}
 contrastive loss, recommendation system, data augmentation, graph convolution network
\end{IEEEkeywords}

\IEEEpeerreviewmaketitle
 
\section{Introduction} 
Recommendation systems  have become an important research field which aims to solve the information overload problem in the information explosion era. The recommendation system  is widely used in many fields, such as e-commerce \cite{ecomrec,alibabaifs }, life services\cite{ Traveltmm,wyxPOI},  social networks\cite{Friendtmm, zhaosocial2 }, entertainment\cite{Micro-Videotmm,zhaoemoji }, and it becomes one of the important technologies in the information age. The top-k   recommendation is the basic problem of recommendation systems which learns the users' preferences through their historical interaction records. Then it recommends top k items to the users that they may like.
 
Deep learning based top-k recommendation algorithms  significantly improve the recommendation performance and become the mainstream research direction in recent years, especially the collaborative filtering based methods.  
These existing algorithms extract advanced semantic features and perform complex feature interactions by employing  MLP\cite{NCF}, CNN\cite{ ONCF}, RNN\cite{DBLP:conf/recsys/DonkersL017}, attention mechanism\cite{AFM,hjmgraph}, etc. The user-item interaction is naturally viewed as a bipartite graph. Graph convolutional networks (GCN) based methods are increasingly integrated with recommendation systems, such as NGCF\cite{NGCF}, LR-GCCF\cite{LRGCCF}, LightGCN\cite{LightGCN}, DGCF\cite{DGCF}. GCN based methods  aggregate features of neighbors as well as higher-order neighbors to obtain  better feature representations of users and items and the performance has been further improved. 

In contrast to the rapid development of the recommendation methods, the loss function has rarely been improved. Bayesian Personalized Ranking (BPR) \cite{BPR} is a widely used loss function for the top-k recommendation, which maximizes the distance between positive and negative pairs. 
Recently, the contrastive loss (CL) function has yielded excellent results in several fields under the contrastive learning framework \cite{simCLR,moco,supcl,pcl,BYOL,GraphCL}. In CL, all non-positive samples in the same batch are used as negative samples,  while BPR uses one or several  negative samples by random sampling. They all learn through a contrastive process, so BPR loss can also be seen as a kind of contrastive loss. 
BPR samples one or several true negative samples, while CL directly treats non-positive items within the same training batch as negative samples. CL can obtain a large number of negative samples simply and quickly while BPR utilizes a small number of samples and requires additional sampling time.




\begin{figure}[t] 
\centering
\includegraphics[width=8.6cm]{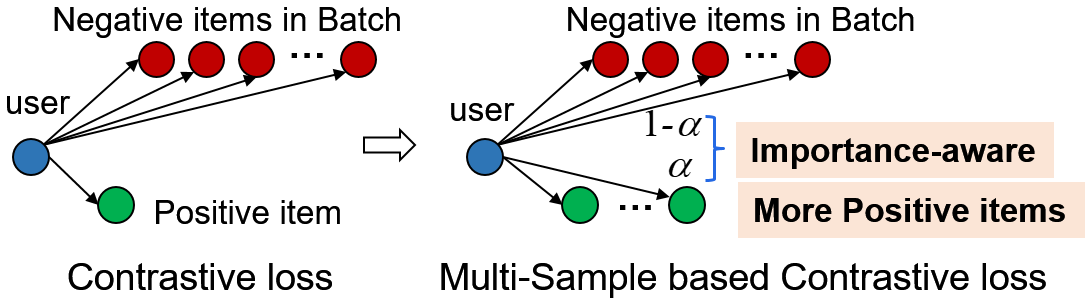} 
\caption{We propose a multi-sample based contrastive loss (MSCL) which distinguishes the importance of positive and negative samples and  makes better use of sparse positive  samples by a new data augmentation method. }  
\end{figure}

However, the importance of  positive and negative samples should be treated differently by CL. (1) CL uses one positive sample and $N$-1 negative samples where $N$ is the batch size, typically 1024, 2048, etc. Thus the imbalance problem or different importance of the one positive sample and negative samples should be tackled. (2) The count of positive samples is very small  thus recommendation systems facing the sparsity problem. Intuitively, the sparser the dataset, the fewer positive samples, and the more important the positive samples should be,  relative to the many negative samples. 
Therefore, positive and negative samples should be treated differently according to the above two reasons. 
 
Another issue we  concern about is the insufficient use of positive samples in the top-k recommendation. As mentioned before, there are very limited positive items of each user in the recommendation system. How to make full use of  the existing  positive samples is a key problem. Data augmentation methods help to solve this problem. Data augmentation methods in the recommendation system  are generally based on graph structures, such as edge or node dropout, masking features, random walk and so on.  The potential of the combined use of positive samples is not exploited.  
 
To solve the above  problems, we propose a new CL based loss function and the basic idea is shown in Fig. 1. For the first problem, we distinguished their different importance by adjusting the weights of positive and negative samples. The hyperparameter $ \alpha$ is the weight of the positive samples which  represents their importance.  To make better use of positive items,  we propose a new  data augmentation method by using multiple positive samples simultaneously.  This data augmentation  makes better use of positive samples for the training space can be expanded because of different combinations of multiple positive samples.  In the original situations, how many items the user has interacted with can be interpreted as how many cases the user can encounter. By a random combination of multiple items, the user can encounter more cases, thus expanding the training space.
Moreover, this data augmentation can be used for many other types of data, not just  graph data.  

In summary, we propose a contrastive loss function based on multiple (positive and negative) samples, which is named as Multi-Sample based Contrastive Loss (MSCL). 
The main contributions of this paper are summarized as follows:
\begin{itemize}
\item We propose a simple but  effective loss function, MSCL, which improves the contrastive loss  to make it suitable for the recommendation system. MSCL  can be applied to many models of recommendation system and is much better than the traditional BPR loss.  
\item The MSCL function distinguishes the importance of positive and negative samples by weighting. It helps to address the imbalance problem of positive and negative samples as well as to enhance the importance of positive samples in sparser datasets. 
\item We propose a new data augmentation method by using multiple positive samples simultaneously which makes better use of the  positive samples. 
\item Experimental results demonstrate the  state-of-the-art performance and many other advantages, such as broad  applicability, high training efficiency. MSCL is  suitable for the top-k recommendation, and it makes the  simple and basic MF more competitive.  
\end{itemize} 

The rest of this paper is organized as follows: In Section II, related works are briefly reviewed. 
To verify the effectiveness of MSCL, we design  sLightGCN\_MSCL which combines MSCL with the best baseline as our method  in Section III. Experiments and discussions  are shown in Section IV. Section V discusses  the advantages of MSCL  with more experiments.
Conclusions are drawn in Section VI.

\section{Related work} 
In this section, we   briefly review related works: the contrastive loss and graph data augmentation methods. Differences between ours and existing works are also presented.
 
 \subsection{ Contrastive Loss (CL) }
   
The contrastive loss has become an excellent tool in unsupervised representation learning. It aims to maximize the similarities of positive pairs and minimize that of negative pairs \cite{simCLR,moco,chopra2005learning, DBLP:conf/cvpr/HadsellCL06,cltmm}. The contrastive loss function is widely used for many kinds of data, such as images, text, audio, graphs, etc. It has  been applied in the field of recommendation\cite{SGL,ContrastiveSequential}. 

Broadly speaking, functions that use pairwise contrastive learning processes are  contrastive loss functions which have many forms.   BPR and triplet loss are  the basic contrast-based and widely used loss functions.  BPR \cite{BPR} loss aims to maximize the distance between positive pair and negative pair, which is proposed for the ranking task and widely used in the top-k recommendation. Triplet loss \cite{schroff2015facenet, DBLP:journals/jmlr/ChechikSSB10, weinberger2006distance} can be used to train samples with small differences, especially for human faces. The samples  are triplets (anchor, positive, negative). The triplet loss is calculated by optimizing the distance between the anchor and positive samples to be smaller than the distance between the anchor and negative samples. 

However, they employ limited pairs of samples. Contrastive loss functions based on multiple pairs of samples are more efficient which contain Multi-class N-pair loss, InfoNCE loss, Non-Parametric Softmax Classifier, NT-Xent loss. 
Multi-class N-pair loss \cite{sohn2016improved} is proposed from a deep metric learning perspective, which greatly improves the triplet loss by jointly pushing out multiple negative samples at each update. InfoNCE loss \cite{oord2018representation, hjelm2018learning, bachman2019learning} is proposed by maximizing a lower bound on mutual information based on Noise-Contrastive Estimation. Non-Parametric Softmax Classifier \cite{Nonpara} is presented by maximizing distinction between instances via a novel non-parametric softmax formulation in an unsupervised feature learning approach. They come from different fields and formula derivation but share a similar form.  NT-Xent loss (the normalized temperature-scaled cross entropy loss) \cite{simCLR, GraphCL} is proposed on these bases, but with the minor difference that the denominator does not contain positive samples. All of them are widely used in contrastive learning framework and always obtain the state-of-the-art results. 

The contrastive loss is used in  the recommendation system in the  contrastive learning framework   for recommendation recently.  SGL\cite{SGL} is proposed for top-k recommendation, CL4SRec \cite{ContrastiveSequential} is used for sequential recommendation. But none of them has improved CL to fit the recommendation field. 

\subsection{Graph Data Augmentation } 

The user-item interaction  records in recommendation systems are naturally viewed as bipartite graphs. Graph-based data augmentation is widely used and studied in graph contrastive learning, which contains both traditional subgraph sampling methods and recently proposed methods \cite{chen2017sampling,GCC,  SGL, GraphCL, MVGCL,GAA}.
User and item are inherently linked and dependent on each other in the user-item bipartite graph.  Data augmentation for  GCN  is also challenging due to  the complex, non-Euclidean structure of the graph, and few works study the data augmentation of graphs. Therefore, graph data augmentation must be tightly integrated with the graph rather than replicating the methods used in  Computer Vision  and Natural Language Processing domains.


Graph data augmentation conforms to the basic assumptions of graph data processing. Node dropping assumes that edge vertex missing does not alter semantics. Edge perturbation is considered to improve the robustness of the semantics against connectivity changes. Masking node features enhance semantic robustness by losing some attributes for each node. Subgraphs assume that local structure can hint the complete semantics \cite{GraphCL}. 

The graph augmentation can be divided into two types, feature-space augmentations and structure-space augmentations: (1) feature-space augmentations are realized by modifying initial node features, such as masking features or adding Gaussian noise, and (2) structure-space augmentations operate on graph structure by adding or removing nodes or edges (edge perturbation),  sub-sampling or subgraphs by random walk, or generating different views using shortest distances or diffusion matrices \cite{MVGCL}. 

Recently, Zhu  et al. \cite{GAA} proposes adaptive graph augmentation  to design augmentation schemes that tend to keep important structures and attributes unchanged while the unimportant links and features are perturbed. Zhao et al.  \cite{Gaug} utilizes a neural edge predictor to predict likely edges for graph augmentation to improve node classification performance.  For top-k recommendations, the latest work, SGL \cite{SGL}, uses three operators on the graph structure, namely node dropout, edge dropout and random walk. And experimental results show that the edge dropout performs the best.
 
\subsection{ Differences with Existing Works } 
Differences with existing CL functions: Many works are done just using  the CL function in the contrastive Learning framework. SGL in the recommendation system employs a multi-tasking mechanism with joint use of CL and BPR.  Despite some improvements are proposed on CL, such as soft contrastive loss \cite{softcl}, debiased contrastive loss \cite{DebiasedCL},  they all  treat  the  weights  of  the  positive  sample  and   negative  samples  as  the  same. And the problem of imbalanced positive and negative samples is still not be concerned. More importantly, how to adapt CL to recommendation systems is a new topic worth investigating, especially to emphasize the importance of positive samples on sparser datasets. Thus, the proposed importance-aware CL is different from the previous works. 

Differences with existing graph data augmentation:  Existing graph data augmentations in the recommendation system  are  common methods in the graph field.  
How to make full use of the limited positive samples to obtain better results, especially for  the recommendation system is an important task and challenge for data augmentation. We randomly sampled a fixed number within one-hop neighbors which is not the same as random dropout or subgraph by random walk on multiple hops. 
More importantly, our  method is a structure-space based augmentation, and traditional structure-space based methods are generally work in the aggregation process of GCNs. 
The traditional augmented data  are used one by one under the same loss which is a serial approach. We use multiple positive samples at the same time and combine them explicitly with the loss function which is a parallel way for better constraints.

\section{Methodology} 

We designed a method named as sLightGCN\_MSCL which combine MSCL with a strong baseline as shown in Fig. 2. to verify the effectiveness of the proposed loss function.
It should be noted that our approach is model-agnostic and can be applied to many methods of recommendation systems. In this section, we first briefly describe the basic methods LightGCN, then  focus on the MSCL function, and finally  analyze the time complexity. 

\begin{figure*}[ht] 
\centering
\includegraphics[width=16cm]{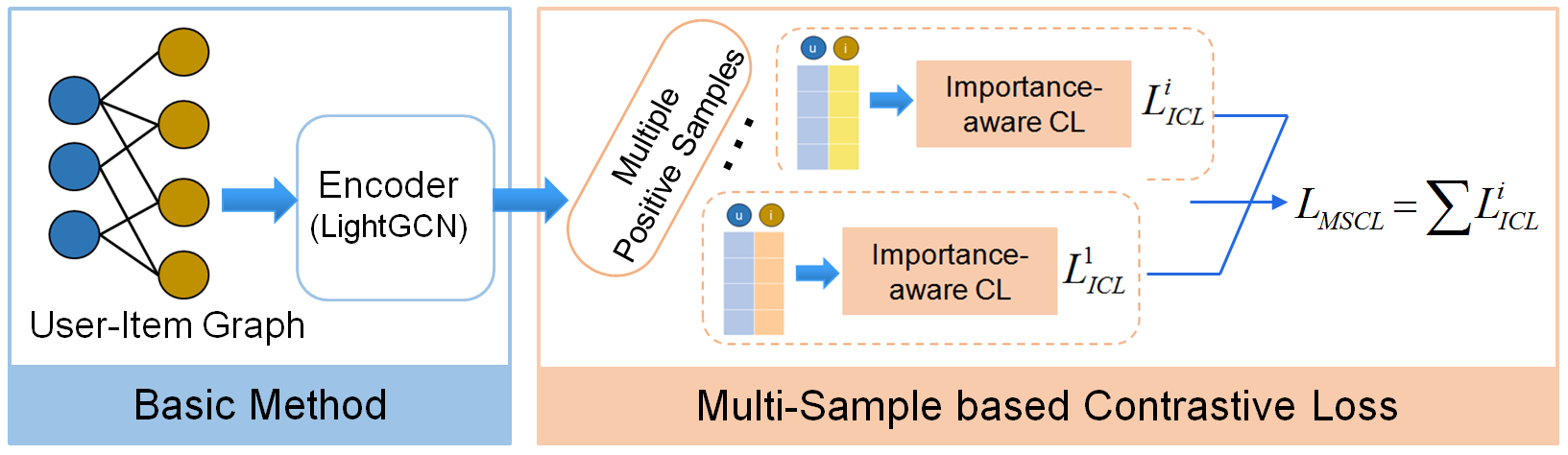}
\caption{ An illustration of our method. Many models can be used as the encoder, LightGCN is used here as an example. } 
\end{figure*}

\subsection{Basic Method }
Recently, graph related methods have shown excellent performances, which treat user-item interactions as graph structures and  adopt graph convolution network. Combined with collaborative filtering, NGCF \cite{NGCF}, LR-GCCF \cite{LRGCCF}, LightGCN \cite{LightGCN}, etc. are excellent models   for top-k recommendation. 
 
LightGCN is the state-of-the-art  method  and is introduced here as our main baseline.  
This model includes only the most essential component in GCN - neighborhood aggregation - for collaborative filtering which is much easier to implement and train and gains substantial improvements. Then neighborhood aggregation is defined as follows: 

\begin{equation}
\boldsymbol{e}_{u}^{(k+1)}=\sum_{i \in \mathcal{N}_{u}} \frac{1}{\sqrt{\left|\mathcal{N}_{u}\right|} \sqrt{\left|\mathcal{N}_{i}\right|}} \boldsymbol{e}_{i}^{(k)}
\end{equation}
\begin{equation} 
\boldsymbol{e}_i^{\left(k+1\right)}=\sum_{u\in \mathcal{N}_i}\frac{1}{\sqrt{|\mathcal{N}_i|}\sqrt{|\mathcal{N}_u|}}\boldsymbol{e}_u^{\left(k\right)}. 
\end{equation}
where  $u, i$ denote the user and the item in the user-item graph, ${e}_u^{\left(k\right)}, {e}_i^{\left(k\right)}$ respectively denote embeddings of $u, i$ of the $k$-th layer. Specially, $k=0$ represents the initialized latent vector; $\mathcal{N}_{u}$ and $\mathcal{N}_{i}$ represent the set of the neighbors of target $u$ and $i$, relatively. The final embedings of users and items are: 
\begin{equation}
\boldsymbol{e}_u=\sum_{k=0}^{K}\alpha_k\boldsymbol{e}_u^{\left(k\right)} 
\end{equation} 
\begin{equation} \boldsymbol{e}_i=\sum_{k=0}^{K}\alpha_k\boldsymbol{e}_i^{\left(k\right)} 
\end{equation}
where  $K $ is the numbers of layers; $ \alpha_k $ denotes the importance of the $k$-th layer embedding, and they can be  treated as a hyper-parameter to be tuned manually, or as a model parameter to be optimized automatically. Following the original paper of LightGCN,  the mean of embeddings from all layers are adopted as the final embeddings, that is $\alpha_k$=1/($K$+1). 

LightGCN-single, a variant of LightGCN are also proposed in the paper, where only the $k$-th embeddings, $ {e}_{u}^{(k)}, {e}_{i}^{(k)} $, are used as final embeddings. This variant, instead of the original LightGCN, is used here for its better performance and is named \textbf{sLightGCN} for short. 
   

The BPR loss is used for training in LightGCN.  We present it here for comparison with MSCL.
\begin{equation}
 L_{BPR}=\sum_{(u, i, j) \in O}-\log \sigma\left(\hat{y}_{u i}-\hat{y}_{u j}\right)
\end{equation}
where $\sigma (\cdot)$ is the logistic sigmoid function.

\subsection{Multi-Sample based Contrastive Loss (MSCL) } 
\subsubsection{The Basic Contrastive Loss (CL) }
Refer to some recent works \cite{simCLR, GraphCL}, we use the NT-Xent  as the original contrastive loss function and then adapt it to the recommendation field. 

The NT-Xent is:
\begin{equation} 
L=-\log \frac{\exp \left(\operatorname{sim}\left(\boldsymbol{z}_{i}, \boldsymbol{z}_{j}\right) / \tau\right)}{\sum_{k=1, k\neq i}^{N} \exp \left(\operatorname{sim}\left(\boldsymbol{z}_{i}, \boldsymbol{z}_{k}\right) / \tau\right)} 
\end{equation}
where $sim \left(\boldsymbol{z}_{i}, \boldsymbol{z}_{j}\right)=\boldsymbol{z}_{i}^{\top} \boldsymbol{z}_{j} /\left\|\boldsymbol{z}_{i}\right\| \| \boldsymbol{z}_{j}\|$, and $z_i$, $z_j$ means the embeddings of  sample $i, j$  in a minibatch, $N$ is the batch size, $\tau$ denotes the temperature parameter. To fit the recommendation domain, we  rewrite it as  $L_{CL}$:
\begin{equation} 
f \left(u, i\right)=\boldsymbol{e}_{u}^{\top} \boldsymbol{e}_{i} /\left\|\boldsymbol{e}_{u}\right\| \| \boldsymbol{e}_{i}\| 
\end{equation} 
\begin{align}
L_{CL} &=-\frac{1}{N}\sum_{(u, i)\in D}\log \frac{\exp \left( f(u, i^+)/\tau\right) }{\sum_{i\in I^-} \exp \left( f(u, i)/\tau\right)} \nonumber \\
&=-\frac{1}{N}\sum_{(u, i)\in D}\Big( f(u, i^+)/\tau - \log\sum_{i\in I^-} \exp \left( f(u, i)/\tau\right))\Big)
\end{align} 
where $D=\left\{(u, i), u\in U, i\in I\right\}$ is a training batch; $U$, $I$ is the set of users and items, respectively; $i^+$ means the positive sample of target user $u$, and $I^-$ is the set of negative samples. $ f \left(u, i\right)$ is the cosine similarity of the $(u, i)$ pair based on their embeddings. We follow the sampling strategy used in \cite{simCLR, GraphCL} that the other non-positive samples in the same batch are seen as negative samples.

\subsubsection{Importance-aware CL (ICL) }
In the basic contrastive loss, the minus sign is preceded by one positive sample, followed by the sum of $N-1$ negative samples which results in imbalance problems. 
In addition, emphasizing the importance of positive samples on sparser datasets is also a problem we want to address. These two problems can be solved together by weighting, an effective and common practice. The importance of positive and negative samples can be adjusted which helps to better backpropagate and make the training more effective. We name the modified CL as importance-aware CL, which is:
\begin{align}
L_{ICL} =& -\frac{1}{N}\sum_{(u, i)\in D}\Big(\alpha f(u, i^+)/\tau \nonumber \\
& -(1-\alpha) \log\sum_{i\in I^-} \exp \left( f(u, i)/\tau\right))\Big)
\end{align} 
where $\alpha $ is a hyperparameter    and $\alpha \in[0, 1]$. 

When  $\alpha=0.5$, the $L_{ICL}$  is the same  as $L_{CL}$. Because problems in the analysis are inevitable, $\alpha=0.5$ is not optimal in general. When $\alpha>0.5$, it means that $N$-1 negative samples need more weights and relatively more losses to be optimized.
When $\alpha<0.5$, it means that the positive samples need to be given more attention, and it may be because there are too few positive items for users in  the recommendation system.  The weighting method is simple yet effective in adapting to a variety of datasets.

\subsubsection{Multiple Positive Samples based CL (MCL)}

To address the problem that positive samples are  insufficiently used, we propose a new data augmentation method that uses multiple positive samples simultaneously. We propose a multi-path based method to use multiple positive samples under the supervision of CL function. The conventional approach is to use a random batch of data and a loss function to form a learning path after the model is computed. We extend this idea by randomly sampling $M$ positive samples to form $M$ paths.  The target user is optimized by $M$ positive samples. The final loss is the sum of $M$ loss functions which is calculated simply and effectively. Thus, the multiple positive samples based CL is:
\begin{equation} 
L_{MCL}=\sum_{m=1}^{M} L_{CL}^{m}
\end{equation} 
where $M$ is  a hyperparameter which is the  number of used positive samples. 

This data augmentation  comes with many benefits. (1)
We keep the same training process of MSCL and the original one, but the difference is theof samples used for each training. 
Suppose the user has $L$ positive samples, there are $C_{L}^{1}$  (Combination formula) possible cases by random sampling for the user at each training time in the original way.  MSCL uses $M$ positive samples simultaneously for the user, so there are  $C_{L}^{M}$ possible cases for each training.  Therefore, the $M$  positive samples  greatly increase  the cases that users can encounter. This  makes augmentation and better usage of the existing positive items. (2) Positive and negative samples form a comparison, thus the expanded positive samples also enlarge the comparable cases. (3) Furthermore, we integrate  this augmentation  with the loss function explicitly in  a parallel way to facilitate more and better constraint and backpropagation for the user.  (4) And this data augmentation method  can be widely applied to graph data as well as various other types of data.

\subsubsection{ Multi-Sample based CL (MSCL)}
We have elaborated on our two improvements,  ICL and MCL.
These two improvements are proposed from different perspectives. Combining them together can solve the  two problems for the top-k recommendation. 
In this case, multiple positive samples and many negative samples are used at the same time.  Therefore, we term it as Multi-Sample based Contrastive Loss, which is defined as follows:
\begin{equation} 
L_{MSCL}=\sum_{m=1}^{M} L_{ICL}^{m}
\end{equation}  
 
Their combination forms a logic for this paper: using multiple (positive  and  negative) samples and solve the problems that exist in them. The proposed function is simple and effective.   Two hyperparameters are introduced, but they are easy to tune. 
 



\subsection{ Model Prediction }
The model prediction is defined as the inner product of the user and item final embeddings: 
 \begin{equation}
\hat{y}_{u i}=\boldsymbol{e}_{u}^{T} \boldsymbol{e}_{i}
\end{equation} 
Based on this prediction, the top-k most similar items are recommended to the user.

The proposed MSCL is used for model training, and the method is named as sLightGCN\_MSCL. MSCL replaces the BPR loss which is used in the original LightGCN. 
Except for the proposed loss function, our method remains the same as the LightGCN.  The L2 regularization for all parameters is also used following LightGCN, and it is omitted here for clarity.
 
\subsection{Complexity Analyses  } 
In this subsection, we analyze the complexity of sLightGCN\_MSCL following SGL \cite{SGL}. Since sLightGCN\_MSCL does not introduce trainable parameters and there is no change of model prediction, the spatial complexity and the time complexity of the model inference are the same as  LightGCN. The complexity of sLightGCN\_MSCL  can be divided into two parts, that of sLightGCN  and  MSCL, and they  are $O(2|E|)+ O(2|E| L d s |E|/N)$,   $O( m N |E| d s )$, where $E $ is the edge in the user-item interaction graph, $ L, s, m $ denote  the number of GCN layers, the number of epochs,  the number of multiple positive samples, and  $ d, N $ denote the embedding size, the batch size, respectively.    For comparison, that of the BPR loss is $O(2|E| d s)$.


 
In fact, the overall amount of calculation is significantly reduced because the number of  training epoch is substantially reduced due to better convergence performance as shown in training efficiency in section V.  MSCL is $O( m N/2 )$ times larger than  the computational cost of BPR, but this is a simple inner product  which is directly accelerated by matrix operations through the GPU.  
Therefore, there is no significant increase in training time  in each epoch as can be seen in  section V.
    
\section{Experiments} 
We first introduce the basic information related to the experiments, such as datasets, evaluation metrics, and hyper-parameter settings. sLightGCN\_MSCL is compared with many strong baselines.  We conduct  ablation studies to verify the effectiveness of the proposed improvements. The main hyper-parameters of sLightGCN\_MSCL are discussed in detail.

\subsection{Datasets} 
To evaluate the effectiveness of MSCL, we conduct experiments on three benchmark datasets: Yelp2018 \cite{ LightGCN, SGL}, Amazon-Book \cite{LightGCN, SGL}, and Alibaba-iFashion \cite{SGL, alibabaifs}. 

Yelp2018: Yelp2018 is adopted from the 2018 edition of the Yelp challenge. The local businesses like restaurants and bars are viewed as the items. 

Amazon-book: Amazon-review is a widely used dataset for product recommendation and Amazon-book from the collection is selected. 

Alibaba-iFashion: Alibaba-iFashion is a large and rich dataset for Fashion Outfit recommendation. 300k users and all their interactions over the fashion outfits are randomly sampled by SGL \cite{SGL}. It is quite sparse, which is a significant  difference from the first two datasets. 

\begin{table} [!h]
 \centering 
 \caption{Statistics of the datasets. } 
 \begin{tabular}{ccccc}
 \toprule
 Dataset & Users & Items & Interactions & Density \\
 \midrule
 Yelp2018 & 31,668 & 38,048 & 1,561,406 & 0.00130 \\
 Amazon-Book & 52,643 & 91,599 & 2,984,108 & 0.00062 \\
 Alibaba-iFashion & 300,000 & 81,614 & 1,607,813 & 0.00007\\
 \bottomrule
 \end{tabular}%
 \label{tab:addlabel0}%
\end{table}
 Three datasets vary significantly in domains, size, and sparsity. 
The statics of the processed datasets are summarized in Table I.  For comparison purposes, we directly use the split data provided in SGL\cite{SGL}.

\begin{table*}[ht] 
 \centering
 \caption{ Overall Performance Comparison }
 \setlength{\tabcolsep}{5mm}
\begin{tabular}{ccccccc} 
 \toprule

\multirow{2}[4]{*}{Method} & \multicolumn{2}{c}{Yelp2018 } & \multicolumn{2}{c}{Amazon-Book } & \multicolumn{2}{c}{Alibaba-iFashion} \\
\cmidrule{2-7}  & Recall & NDCG & Recall & NDCG & Recall & NDCG \\
 \midrule
 MF & 0.0441 & 0.0353 & 0.0329 & 0.0249 & 0.1020 & 	0.0474\\
 \midrule
NGCF &  0.0579  &  0.0477  &  0.0344  &  0.0263  &  0.1043  &  0.0486  \\
LR-GCCF & 0.0591 & 0.0485 & 0.0378 & 0.0292 & 0.1110 & 0.0529 \\
LightGCN &   0.0639  &   0.0525  &  0.0411  &  0.0315  &  0.1078  &  0.0507  \\
sLightGCN & 0.0649 & 	0.0525 & 	0.0469 & 	0.0363 & 	\underline{0.1160} & 	\underline{0.0553} \\ 
\midrule
Mult-VAE &  0.0584  &  0.0450  &  0.0407  &  0.0315  &  0.1041  &  0.0497  \\ 
SGL &  \underline{0.0675}  &  \underline{0.0555}  &  \underline{0.0478}  &  \underline{0.0379}  &  0.1126 &  0.0538 \\
\midrule
LightGCN\_MSCL(ours) &  0.0681  &  0.0564  &  0.0500  &  0.0391  &  0.1144  &  0.0546  \\
sLightGCN\_MSCL(ours) & \textbf{0.0691} & \textbf{0.0568} & \textbf{0.0580} & \textbf{0.0466} & \textbf{0.1201} & \textbf{0.0578} \\
 \bottomrule
\end{tabular} 
\end{table*}

\subsection{Evaluation Metrics } Following NGCF, LightGCN, and SGL \cite{NGCF, LightGCN, SGL}, two widely used evaluation metrics, Recall@K and NDCG@K where K=20, are used to evaluate the performance of top-k recommendation. Recall measures the number of items that the user likes in the test data that has been successfully predicted in the top-k ranking list. NDCG considers the  positions of the items and higher scores are given if the items are ranked higher. It is a metric about ranking and thus is  important for the top-k recommendation. The larger the values, the better the performance for both metrics. 
 
\subsection{Hyper-parameter Settings} 
We implement our proposed method on top of the official code of LightGCN\footnote{https://github. com/gusye1234/LightGCN-PyTorch} based with Pytorch. We replace the loss function and follow LightGCN's settings as much as possible. The embedding size is fixed to 64 and the default batch size is 2048 for all models. The learning rate and L2 regularization coefficients are chosen by grid search in the range of $\{0. 0001, 0. 001, 0. 01\}$ and $\{1e-5, 1e-4, \cdots, 1e-2\}$.  These are hyper-parameters of the original LightGCN.  We adjust hyper-parameters of MSCL, $M$ and $\tau$, in the ranges $\{1, 3, 5, \cdots, 15\}$, $\{0. 1, 0. 2, 0. 5, 1. 0\}$, respectively. And $\tau$ is 0.1 or 0.2 usually. The weight $\alpha$ is adjust in [0.4,0.7]. 

\subsection{Compared Methods}

To demonstrate the  performance of our method, we select many  strong baselines for comparison.  NGCF \cite{NGCF}, LR-GCCF \cite{LRGCCF}, LightGCN \cite{LightGCN} are the competing baselines with GCN of top-k recommendation recently which having shown to outperform several methods including GC-MC \cite{GCMC}, PinSage \cite{PinSage}, NeuMF \cite{NCF} since the previous works \cite{NGCF, LightGCN, LRGCCF}. The latest methods, SGL \cite{SGL}, is also selected which is a self-supervised based method. In addition, the basic method and the variable autoencoder-based methods, MF and Mult-VAE, are compared.

MF: This is a traditional method based on matrix factorization which is based only on the embeddings of users and items, namely ${e}_{u} $ and $ {e}_{i}$. 

NGCF \cite{NGCF}: 
NGCF integrates the bipartite graph structure into the embedding process based on the graph convolutional network. It explicitly exploits the collaborative signal in the form of high-order connectivities by propagating embeddings on the graph structure. 
  
LR-GCCF \cite{LRGCCF}: This method enhances the recommendation performance with less complexity by removing the non-linearity. The final embeddings are the same as NGCF. 

LightGCN and sLightGCN \cite{LightGCN}: LightGCN is the state-of-the-art GCN based collaborative filtering model, and sLightGCN is a variant. They are described in detail in Section III. 

Mult-VAE \cite{multiVAE}: Mult-VAE  extends variational autoencoders (VAEs) to collaborative filtering and uses a multinomial likelihood for the data distribution. Besides, it introduces an additional regularization parameter for optimization. It can be seen as a special case of self-supervised learning(SSL) for recommendation. 

SGL \cite{SGL}: SGL is the latest baseline for top-k recommendations. It introduces self-supervised learning into the recommendation system based on the contrastive learning framework. It is implemented on LightGCN and uses a multi-task approach that unites the contrastive loss and the BPR loss function. SGL mainly benefits from graph contrastive learning to reinforce user and item representations. Following the paper, the edge drop based SGL achieving the best performance are adopted here.

\begin{table*}[htb]
 \centering
 \caption{Ablation study }
 \setlength{\tabcolsep}{3mm}
 \begin{tabular}{ccccccccc}
 \toprule
 \multirow{2}[4]{*}{Loss} & \multicolumn{1}{c}{\multirow{2}[4]{*}{Importance-aware}} & \multicolumn{1}{c}{\multirow{2}[4]{*}{Multi-positive samples}} & \multicolumn{2}{c}{Yelp2018 } & \multicolumn{2}{c}{Amazon-Book } & \multicolumn{2}{c}{Alibaba-iFashion} \\
\cmidrule{4-9}  & & & Recall & NDCG & Recall & NDCG & Recall & NDCG \\
 \midrule
 CL & & & 0.0655 & 0.0541 & 0.0480 & 0.0399 & 0.1152 & 0.0556 \\
 ICL & \checkmark & & 0.0668 & 0.0548 & 0.0544 & 0.0437 & 0.1165 & 0.0558 \\
 MCL & & \checkmark & 0.0677 & 0.0559 & 0.0516 & 0.0425 & 0.1184 & 0.0573 \\
 MSCL & \checkmark & \checkmark & 0.0691 & 0.0568 & 0.0580 & 0.0466 & 0.1201 & 0.0578 \\
 \bottomrule
 \end{tabular}%
 \label{tab:addlabel}%
\end{table*}%

\subsection{Performance Comparison} 

The performance comparison  on the three datasets is shown in Table II.  The best results are shown in bold while underlined scores are the second best. We follow the experimental results of SGL \cite{SGL}, except for MF, LR-GCCF, and our methods. After statistical analysis, the standard deviations on Recall  and NDCG are not big than $ \pm 0.0002$ under different initialization seeds.  We have the following observations: 

MF is the most basic and simplest method and performs the worst. 
NGCF, LR-GCCF, and LightGCN are GCN based methods. 
NGCF achieved  improvements relative to MF by  introducing the GCN method into top-k recommendations, especially  on the Yelp2008 dataset.
 LR-GCCF, LightGCN and sLightGCN can be seen as improvements of NGCF.
Their performances are better than NGCF, and these results are consistent with the performance in the original paper.   These three methods show the significant role of graph convolution methods in recommendation systems. LightGCN is the strongest baseline and becomes the basis for subsequent methods, such as SGL and our method.  LightGCN removes the nonlinear activation layer and learning parameters making the model more applicable to recommendation systems rather than simply employing GCN which illustrates that the GCN method should be modified to fit the recommendation system. 

Mult-VAE and SGL are methods that belong to self-supervised learning (SSL). The results of Mult-VAE are generally better than NGCF, indicating  that  the variational auto-encoder based method and self-supervised learning is competitive for recommendation. 
The results of SGL show that it has a clear boost compared with LightGCN and suboptimal results are obtained on two datasets which demonstrates the advancement of contrastive learning methods.  

The proposed sLightGCN\_MSCL is the best among all methods. LightGCN\_MSCL   is also listed as a variant of our method, which is also superior to other methods. Compared to the latest and best method SGL, the improvements of sLightGCN\_MSCL on Yelp2018, Amazon-Book, and Alibaba-iFashion are 2.37\%, 21.34\%, 6.67\% on Recall, 2.34\%, 22.96\%, 7.43\% on NDCG, respectively. SGL uses CL and BPR jointly in the multi-task learning approach without exploiting the potential of CL. Our approach is simpler and consumes less time which can be seen in Training Efficiency in Section V. This shows the correctness of improving CL.
 

\subsection{ Ablation Study} 
MSCL combines two components,  the different importance of positive and negative samples and the use of multiple positive samples. ICL, MCL denote importance-aware CL and multiple positive samples based CL respectively. They are shown in Equations (8)-(10).
  
\subsubsection {The effectiveness of the two components}
Detailed ablation studies demonstrate the effectiveness of our two components  as shown in Table III.  The comparison of CL and ICL, MCL and MSCL show the effectiveness of adding weights to distinguish the importance of positive and negative samples. The comparison of CL and MCL, ICL and MSCL illustrate the effectiveness of data augmentation based on multiple positive samples. All the results, the two evaluation metrics on three datasets in these four comparison groups, in Table III  consistently demonstrate the effectiveness of the two components. 
  
Besides, we find the three datasets perform differently on the two components. Amazon-Book benefits more from adding weights to distinguish the importance while the other two datasets improve more significantly on multiple positive samples. This shows that the proposed two components are effective, but perform differently depending on the dataset. 

\subsubsection {Detailed analysis about the role of  multiple positive samples}

\begin{figure*}[ht] 
\centering
\includegraphics[width=16cm]{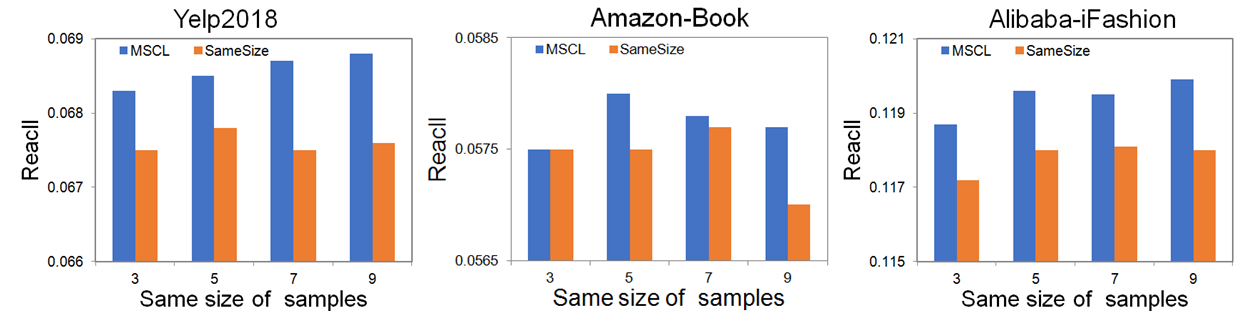}
\caption{ Detailed analysis about the role of  multiple positive samples.  Our data augmentation approach also increases the  count of comparisons in the same epoch.   MSCL still better than the same size of comparisons which indicates MSCL makes better use of  the limited number of positive samples.  }    
\end{figure*}

More comparisons in training tend to yield better results such as a large batch size. Our data augmentation approach of using multiple positive samples also increases the count of comparisons in each epoch.  Therefore, one of the reasons for the good performance  of  multiple positive samples also involves more comparisons. 
However, we want to show that our proposed approach makes better use of positive samples, except for thecountof comparisons. Experiments with the same number of comparisons need to be done  to exclude this factor. We expand the batch size of ICL to $M$*$N$ because a user is compared with $M$*$N$ items in MSCL, where $N$ is the training batch size.

The results are shown in Fig. 3. Our method consistently outperforms the latter on the three datasets.  Overall, the performance of the same batch size peaks and falls back  as batch size increases, especially on Amazon-Book. And they are significantly worse than the performances of multiple positive samples when $M$=9. Yelp2018 and  Alibaba-iFashion  have the same trend of change in Fig. 3  which is different from that of Amazon-Book. 
This is consistent with the above observation in Table III that Yelp2018 and  Alibaba-iFashion behave differently from Amazon-Book.  In summary, the combination of multiple positive samples makes better use of the limited number of positive samples.




\begin{figure*}[ht] 
\centering
\includegraphics[width=16cm ]{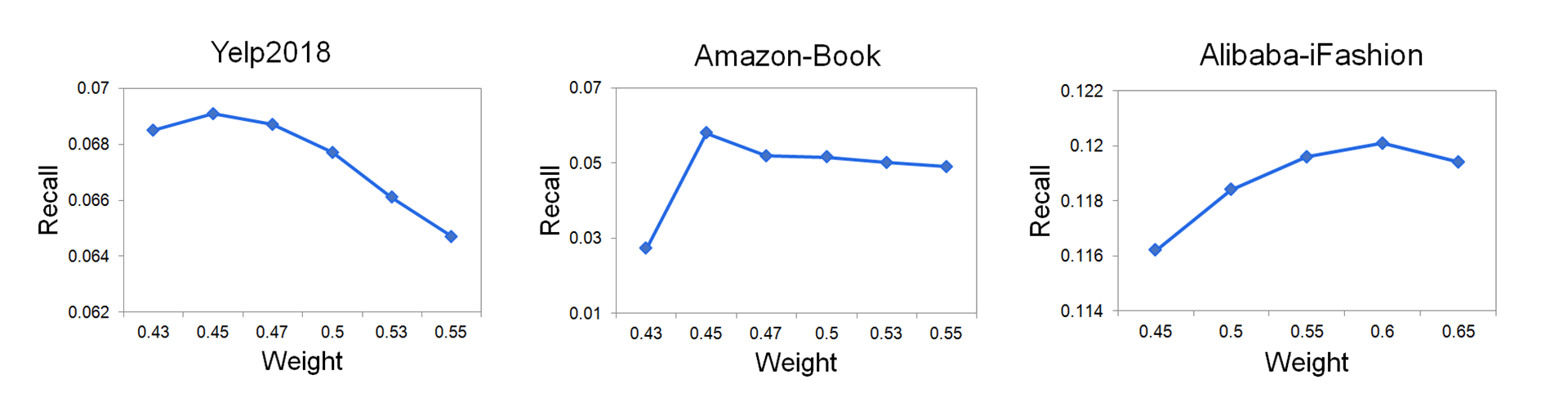}
\caption{Impact of the weight $\alpha$. Different datasets have different optimal weights. Thousands of negative samples in the first two datasets should be given more weight. Alibaba-iFashion is too sparse and thus the positive samples are more important. }
\end{figure*}

\begin{figure*}[ht] 
\centering
\includegraphics[width=16cm]{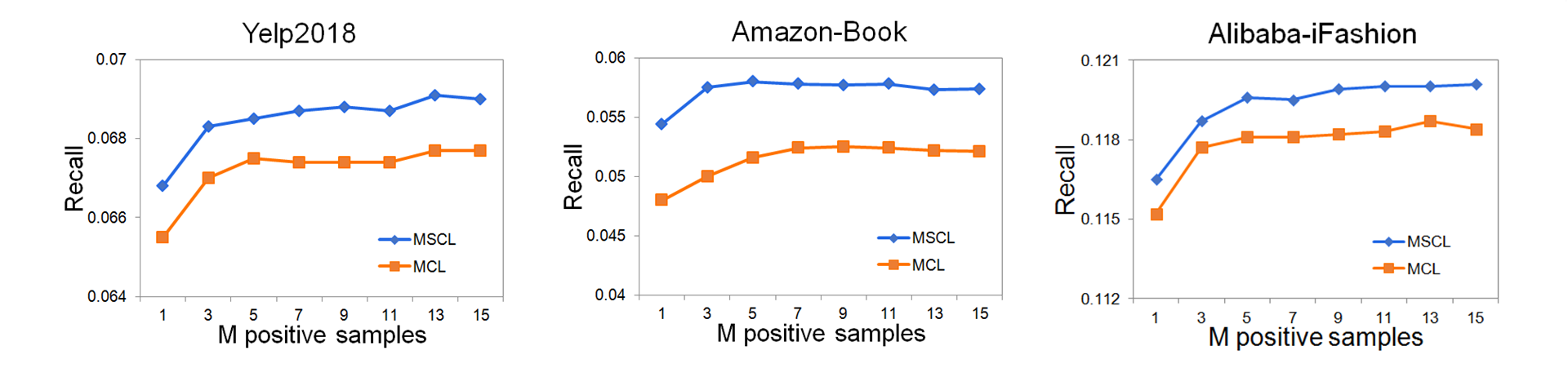}
\caption{Impact of the number of positive samples. The effectiveness of adding positive samples is consistently shown on both curve. }
\end{figure*}

\subsection{Discussion} 

\subsubsection{Impact of the Weight} 

 
We adjust the weight $ \alpha$ and the results are shown in Fig. 4. The results show that both weighting methods achieve better results relative to unweighted when $\alpha$ is 0.5. The first two datasets both obtain the best performance at 0.45, while Alibaba-iFashion reaches the best at 0.60. The main reason for this difference is that Alibaba-iFashion is the sparsest  dataset and has few  positive samples of users.  Each user has 49.3, 56.7, and 6.4 positive items on average on the three datasets, respectively. For Yelp2018 and Amazon-Book, the imbalance problem is the main issue, and thus thousands of negative samples do require relatively more weights to learn better. 
Compared to the other two datasets, positive items of Alibaba-iFashion are so few that positive samples should be more important and given more weight. 
This illustrates that the first two datasets  benefit mainly from solving the imbalance problem and the last dataset  benefits mainly from increasing the importance of a  limited number of positive samples.
It also demonstrates that the weighting approach can solve these two both problems to balance the importance of positive and negative samples and can adapt to different  datasets, despite its simplicity. 

\subsubsection{Impact of the number of positive samples } 
 
Both MSCL and MCL  are able to illustrate the role of multiple positive samples and the results are shown in Fig. 5.  All the results of MSCL and MCL with multiple positive samples are significantly better than those with only one positive sample on the left. So the proposed data augmentation method does make better use of the positive samples.  MSCL and MCL  have the same tend on the three datasets. As the number of positive samples increases, MSCL and MCL start with a significant improvement, and then change flatly. It can be seen from Fig. 5 that about 5 or 7 is appropriate, and more positive samples tend to be slightly better.

Besides, it is consistent with expectations that the results in the left bottom of each dataset in Fig. 5  are the worst  which do not incorporate any improvements. It also shows that all  results of MSCL are better than the MCL with the same number of positive samples, demonstrating the effectiveness  of the importance-aware method.  
 



\section{Advantages of MSCL}   

\begin{table*}[ht] 
 \centering
 \caption{Performance of MSCL compared with BPR on different methods } 
 \begin{tabular}{cllllll}
 \toprule
 \multirow{2}[4]{*}{Method} & \multicolumn{2}{c}{Yelp2018 } & \multicolumn{2}{c}{Amazon-Book } & \multicolumn{2}{c}{Alibaba-iFashion} \\
\cmidrule{2-7}  & Recall & NDCG & Recall & NDCG & Recall & NDCG \\
 \midrule
 MF\_BPR & 0.0441 & 0.0353 & 0.0329 & 0.0249 & 0.1020 & 0.0474 \\
 MF\_MSCL & 0.0657\tiny(48.98\%) & 0.0538\tiny(52.41\%) & 0.0478\tiny(45.29\%) & 0.0369\tiny(48.19\%) & 0.1185\tiny(16.18\%) & 0.0576\tiny(21.52\%) \\
 \midrule
 NGCF\_BPR & 0.0579 & 0.0477 & 0.0344 & 0.0263 & 0.1043 & 0.0486 \\
 NGCF\_MSCL & 0.0655\tiny(13.13\%) & 0.0538\tiny(12.79\%) & 0.0481\tiny(39.83\%) & 0.0375\tiny(42.59\%) & 0.1152\tiny(10.45\%) & 0.0565\tiny(16.26\%) \\
 \midrule
 LR-GCCF\_BPR & 0.0591 & 0.0485 & 0.0378 & 0.0292 & 0.1072 & 0.0507 \\
 LR-GCCF\_MSCL & 0.0658\tiny(11.34\%) & 0.0543\tiny(11.96\%) & 0.0465\tiny(23.02\%) & 0.0360\tiny(23.29\%) & 0.1119\tiny(4.38\%) & 0.0533\tiny(5.13\%) \\
 \midrule
 sLightGCN\_BPR  & 0.0649 & 0.0525 & 0.0469 & 0.0363 & 0.1160 & 0.0553 \\
 sLightGCN\_MSCL & \textbf{0.0691}\tiny(6.47\%) & \textbf{0.0568}\tiny(8.19\%) & \textbf{0.0580}\tiny(23.67\%) & \textbf{0.0466}\tiny(28.37\%) & \textbf{0.1201}\tiny(3.53\%) & \textbf{0.0578}\tiny(4.52\%) \\
 \bottomrule
 \end{tabular}%
 \label{tab:addlabel2}%
\end{table*}%

\begin{figure*}[ht] 
\centering
\includegraphics[width=16cm]{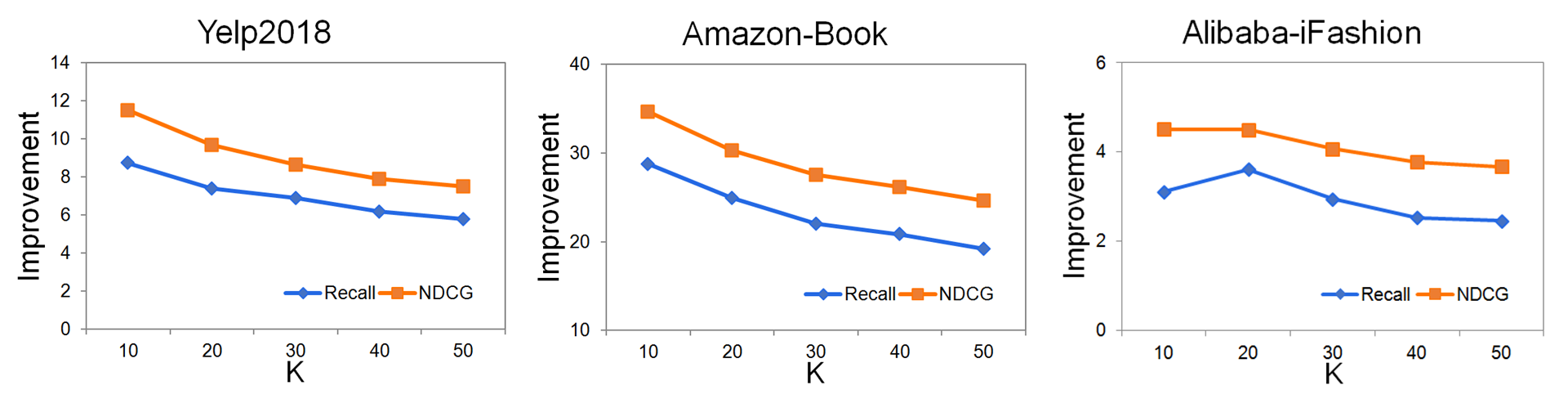}
\caption{Better improvements on NDCG for top-k recommendation. The figure shows the percentage improvement of MSCL over BPR on Recall@K and NDCG@K at different K. Higher improvement of NDCG@K  than Recall@K shows  MSCL is more suitable for top-k recommendation. }
\end{figure*} 

We have obtained optimal results of MSCL by the method  sLightGCN\_MSCL on top-k recommendation. We focus on  the proposed loss function MSCL in this section.
MSCL is simple and easy to implement, but it also has many other advantages, such as applicability, more suitable for the top-k recommendation, high training efficiency. In addition, MSCL improves the simplest and most basic model MF  significantly making it more valuable for applications. Finally, as an extension, we  verify that the problems and the improvements of this paper are also generalizable to multiple samples based BPR function.

\subsection{Applicability of MSCL} To show the applicability  of MSCL, we apply it to many methods and compare it with the  BPR loss, and  methods with these two loss are named as ``*-MSCL", ``*-BPR".  

The results are shown in Table IV, and the percentage of improvements relative to BPR are also presented. MSCL-based methods outperform  the BPR-based methods on all results on the three datasets, and  have significant improvements on  MF, NGCF and LR-GCCF.  sLightGCN\_MSCL consistently obtains the best results on all datasets and has desirable improvements.   In particular,   the improvement on the Amazon-Book dataset is still about 25\%.

Furthermore, MF is the most fundamental method just based on embeddings in the recommendation field, and many methods can be seen as developments of MF. Theoretically, MSCL is suitable for all embedding based methods. Thus, the effectiveness of MF shows that MSCL can be widely used in recommendation systems.  The above experimental results and analysis show that our proposed MSCL is model-agnostic and widely adaptable.
 
\begin{figure*}[ht] 
\centering
\includegraphics[width=16cm]{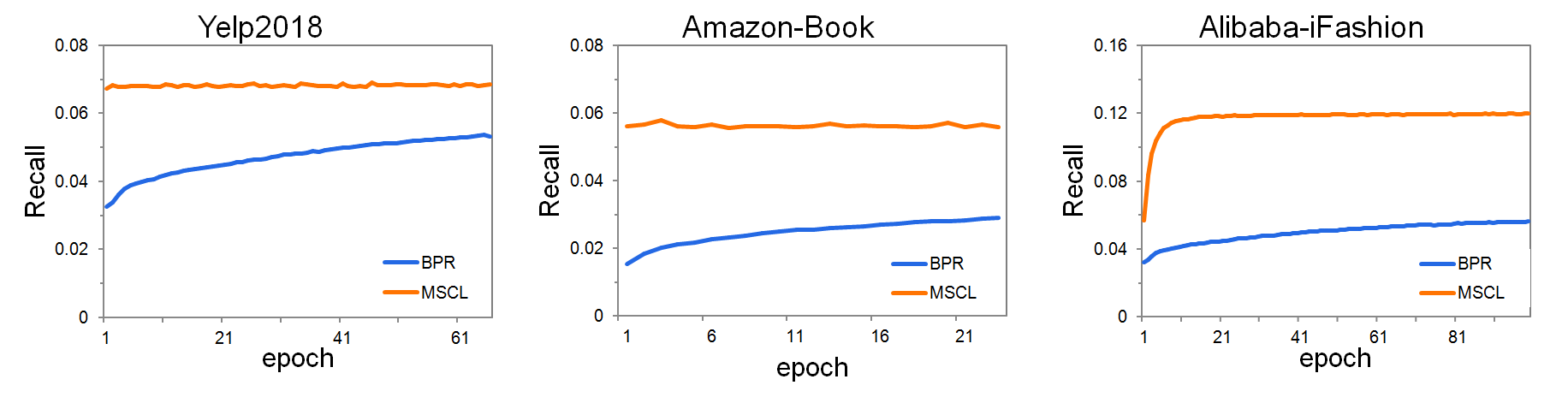}
\caption{ Training Efficiency. Testing recall of MSCL and BPR with sLightGCN on three datasets. Here the  total training epochs of MSCL are shown, the curve of BPR is too long and thus shows the same training epochs as MSCL. } 
\end{figure*}

 
\subsection{ Suitable for the Top-k Recommendation }  
We think that MSCL is more suitable for the Top-k recommendation task. This can be illustrated by theoretical analysis and experimental results.
 
Theoretically, MSCL is compared with the loss function BPR. The common goal of both BPR and MSCL is to learn better feature representation by comparing between positive and negative samples. BPR uses a limited number of comparisons, usually one or several while MSCL employ thousands. Moreover, MSCL improves the quality of comparison by distinguishing the importance of positive and negative samples and makes better use of the few positive samples.  MSCL makes the similarities between positive and negative samples more accurate through more and better comparisons.  The top-k recommendation is a ranking task, and BPR is proposed specifically for ranking tasks.  MSCL outperforms BPR in terms of theoretical and experimental results.  Therefore, MSCL could get better ranking results and makes more sense for top-k  recommendations.    

Experimentally, the improvement of the evaluation metric NDCG  is more obvious. NDCG is a ranking related metric which is more meaningful for ranking and top-k recommendation  task compared to Recall. The following two observations support us well: (1) In Table IV, we found that the NDCG boost generally more than Recall. On average, the improvements are 19.98\%, 32.95\%, 8.64\% on Recall while that are 21.34\%, 35.61\%, 11.86\% on NDCG for the three datasets respectively. This shows the superiority of MSCL for top-k recommendation on different methods. (2) Fig. 6 shows the more significant improvement of MSCL over BPR on NDCG compared to Recall with different $K$. This shows that the ranking performance of NDCG is consistently higher than Recall even as $K$ changes. 
 
\subsection{The Improvement of  MSCL  on MF } The performance of  MF\_MSCL  is particularly noteworthy in Table IV.

(1) MF\_MSCL gains the most significant improvement, among all MSCL-based versus BPR-based methods. The results are even better than those of all BPR-based methods, sLightGCN\_BPR included. It indicates MSCL with the most basic and simple method is significantly better than excellent methods recently proposed, even the state-of-the-art GCN methods. Thus, to some extent, a good loss function works better than  new models.

(2) In addition, we find that the results MF\_MSCL are also  competitive. They are close to or better than that of NGCF\_MSCL and LR-GCCF\_MSCL on all datasets, and are  close to sLightGCN\_MSCL on the Alibaba-iFashion dataset. This indicates that MSCL is also effective in directly optimizing embeddings without a complex model, such as GCN. 

These observations also indicate that  MSCL can achieve competitive results in the simplest baseline, which is also consistent with the latest research Graph-MLP \cite{graphmlp}.  
Graph-MLP indicates that it is sufficient for learning discriminative node representations only by MLP and graph based CL,  without the complex GCN. 
Compared with Graph-MLP, MF\_MSCL is more concise and simple. It is based only on embeddings and improved CL functions, which is still effective even without MLP.  Graph-MLP does not optimize the CL loss function  which is what we do, this shows the great potential of MSCL. 

Three other points need to  be highlighted. (1) As the most basic and simple method, MF\_MSCL can be widely used in various tasks of recommendation systems, not only the top-k tasks. The applicability of MSCL is best illustrated by  MF\_MSCL.   (2) MF\_MSCL also has other advantages of MSCL presented in this section, such as being more suitable for top-k recommendation and fast convergence. 
(3) Importantly, it is valuable for applications with high space and time requirements or industrial applications at a large scale.

\subsection{Training Efficiency}

\begin{table}[!h]
 \centering
 \caption{Actual train time per epoch}
 \begin{tabular}{cccccc}
 \toprule
 \multirow{2}[4]{*}{} & \multirow{2}[4]{*}{BPR} & MSCL & MSCL & MSCL & MSCL \\
\cmidrule{3-6}  & & M=1 & M=5 & M=10 & M=15 \\
 \midrule
 Yelp2018 & 13 & 12 & 15 & 19 & 22 \\
 Amazon-Book & 64 & 61 & 65 & 71 & 77 \\
 Alibaba-iFashion & 17 & 16 & 19 & 22 & 25 \\
 \bottomrule
 \end{tabular}%
 \label{tab:addlabel3}%
\end{table}%

The training efficiency of MSCL is also significantly improved as shown in Fig. 7. Because of the large difference in loss values, following LightGCN, SGL, the test performance on three datasets are used to show the convergence speed. In terms of the number of training epochs required to achieve optimal performance, more than 900 epochs are required for BPR, while  MSCL achieves the best performance at 46, 3, and 90 epochs on the three datasets respectively. BPR requires too many epochs for convergence while MSCL converges earlier, so we  adopt the same number of epoch as MACL for comparison.

For the first two datasets, MSCL converges directly to the high values approximating the final performance with slight fluctuations while BPR  converges slowly at lower values. 
On the third dataset, it is slightly more difficult to converge due to the  sparsity of the dataset. 
MSCL converges sharply by about 5 epochs to the value that approximates the final performance. 
This all shows that MSCL has a fast convergence capability. The training efficiency  is improved at least tens of times on different datasets in terms of train epochs as mentioned before.   The main reason for the high training efficiency is that multiple samples are learned at the same time as demonstrated in \cite{chen2017sampling}. 

Moreover, in terms of actual training time, MSCL does not significantly increase the training time per epoch. Table V shows the average time consumption  in each epoch in seconds. MSCL takes less time than BPR when one positive sample is used as shown in the first two columns of the table. Because BPR requires negative sampling while MSCL needs not and the computation with multiple negative samples is accelerated by the GPU. When the number of positive samples increases by 1, the average time increased on the three data sets is 0.8s. Such time consumption is completely negligible. When $M$=5, MSCL and BPR consume the same time, but the performance is much better than BPR. Relative to the latest SGL \cite{SGL} based on the contrastive learning framework, it takes about 3.7x larger than LightGCN while ours is about 1.5 times of LightGCN. 
 
The above analyses demonstrate that MSCL has remarkable  improvement in convergence speed and training efficiency than BPR. And there is no significant increase in time consumption per epoch, which  is an advantage over SGL in terms of performance and time consumption.

\subsection{Multi-Sample based BPR Loss (MSBPR) }
 
\begin{table}[ht]
 \centering
 \caption{ performance comparison among BPR, MSBPR and MSCL }
 \begin{tabular}{ccccc}
 \toprule
  & & BPR & MSBPR & MSCL \\
 \midrule
 \multirow{2}[2]{*}{Yelp2018 } & Recall & 0.0649 & 0.0670 & 0.0691 \\
  & NDCG & 0.0525 & 0.0552 & 0.0568 \\
 \midrule
 \multirow{2}[2]{*}{Amazon-Book } & Recall & 0.0469 & 0.0458  & 0.058 \\
  & NDCG & 0.0363 & 0.0371 & 0.0466 \\
 \midrule
 \multirow{2}[2]{*}{Alibaba-iFashion} & Recall & 0.1160 & 0.1172 & 0.1201 \\
  & NDCG & 0.0553 & 0.0564 & 0.0578 \\
 \bottomrule
 \end{tabular}%
 \label{tab:addlabel4}%
\end{table}%

The proposed MSCL combines ICL and MCL aiming to solve the problem of different importance of positive and negative samples and insufficient use of positive samples. The problems and solutions are also fit for BPR. Therefore, we modify the loss function of the multi-sample based BPR in the same way and present the MSBPR function. In this case, the same sampling method of MSCL is used by MSBPR. The  formula of MSBPR is as follows:
\begin{equation}
\begin{aligned}
L_{M S B P R} &=\sum_{m=1}^{M} \sum_{(u, i) \in D}-\log \sigma\left(\alpha f\left(u, i^{+}\right) / \tau\right.\\
&\left.-(1-\alpha) f\left(u, i^{-}\right) / \tau\right)
\end{aligned}
\end{equation}
where $\sigma$ is the logistic sigmoid. We use $ f(u, i) $ instead of $ \hat{y}_{ u i}$ by drawing on comparative learning because the $ \hat{y}_{ u i}$ based approach doesn't work. 

The results are shown in Table VI, and the baseline is sLightGCN. With the only exception in all results that the recall of MSBPR is worse than BPR on Amazon-Book, the overall MSBPR-based methods are better than BPR. It shows that our proposed idea can be extended to BPR and other pair-wise based loss functions.   In addition, we find that MSCL works better than MSBPR, especially on the Amazon-Book dataset, which shows the superiority of the CL function again and the correctness of improving CL in this paper. 
Therefore, MSCL is better than BPR and MSBPR  in the field of recommendation systems. 

\section{Conclusion} 
In this paper, we propose MSCL function for the multi-sample based recommendation systems. We distinguish the different importance of positive and negative samples and propose a new data augmentation method to make better use of positive samples. MSCL is  simple but obtains optimal results. More importantly, it has the advantages of wide applicability to various models, suitability for the top-k recommendation, and high training efficiency.   MSCL makes simple and basic MF more valuable for industrial applications. These advantages make MSCL more competitive for top-k recommendation tasks.

This work represents an initial attempt to exploit improved CL for the recommendation. We believe that other improvements based on CL are an important direction. 
The two problems, the different importance of positive and negative samples, insufficient use of positive samples, are still valuable and deserve to be studied in depth. The proposed MSCL has the potential to be extended  to graph-related fields as well as other fields.

\ifCLASSOPTIONcaptionsoff
 \newpage
\fi


\bibliographystyle{IEEEtran} 


\begin{thebibliography}{10}

\bibitem{ecomrec}
Sanshi Yu, Zhuoxuan Jiang, Dongdong Chen, Shanshan Feng, Dongsheng Li, Qi~Liu,
  and Jinfeng Yi.
\newblock Leveraging tripartite interaction information from live stream
  e-commerce for improving product recommendation.
\newblock In {\em {KDD}}, pages 3886--3894. {ACM}, 2021.

\bibitem{alibabaifs}
Wen Chen, Pipei Huang, Jiaming Xu, Xin Guo, Cheng Guo, Fei Sun, Chao Li,
  Andreas Pfadler, Huan Zhao, and Binqiang Zhao.
\newblock {POG:} personalized outfit generation for fashion recommendation at
  alibaba ifashion.
\newblock In {\em {KDD}}, pages 2662--2670. {ACM}, 2019.

\bibitem{Traveltmm}
Zhenxing Xu, Ling Chen, Yimeng Dai, and Gencai Chen.
\newblock A dynamic topic model and matrix factorization-based travel
  recommendation method exploiting ubiquitous data.
\newblock {\em IEEE Transactions on Multimedia}, 19(8):1933--1945, 2017.

\bibitem{wyxPOI}
Yuxia Wu, Ke~Li, Guoshuai Zhao, and Xueming QIAN.
\newblock Personalized long- and short-term preference learning for next poi
  recommendation.
\newblock {\em IEEE Transactions on Knowledge and Data Engineering}, pages
  1--1, 2020.

\bibitem{Friendtmm}
Shangrong Huang, Jian Zhang, Lei Wang, and Xian{-}Sheng Hua.
\newblock Social friend recommendation based on multiple network correlation.
\newblock {\em IEEE Transactions on Multimedia}, 18(2):287--299, 2016.

\bibitem{zhaosocial2}
Guoshuai Zhao, Xiaojiang Lei, Xueming Qian, and Tao Mei.
\newblock Exploring users' internal influence from reviews for social
  recommendation.
\newblock {\em IEEE Transactions on Multimedia}, 21(3):771--781, 2019.

\bibitem{Micro-Videotmm}
Xusong Chen, Dong Liu, Zhiwei Xiong, and Zheng{-}Jun Zha.
\newblock Learning and fusing multiple user interest representations for
  micro-video and movie recommendations.
\newblock {\em IEEE Transactions on Multimedia}, 23:484--496, 2021.

\bibitem{zhaoemoji}
Guoshuai Zhao, Zhidan Liu, Yulu Chao, and Xueming Qian.
\newblock Caper: Context-aware personalized emoji recommendation.
\newblock {\em IEEE Transactions on Knowledge and Data Engineering},
  33(9):3160--3172, 2021.

\bibitem{NCF}
Xiangnan He, Lizi Liao, Hanwang Zhang, Liqiang Nie, Xia Hu, and Tat{-}Seng
  Chua.
\newblock Neural collaborative filtering.
\newblock In {\em {WWW}}, pages 173--182. {ACM}, 2017.

\bibitem{ONCF}
Xiangnan He, Xiaoyu Du, Xiang Wang, Feng Tian, Jinhui Tang, and Tat{-}Seng
  Chua.
\newblock Outer product-based neural collaborative filtering.
\newblock In {\em {IJCAI}}, pages 2227--2233. ijcai.org, 2018.

\bibitem{DBLP:conf/recsys/DonkersL017}
Tim Donkers, Benedikt Loepp, and J{\"{u}}rgen Ziegler.
\newblock Sequential user-based recurrent neural network recommendations.
\newblock In {\em RecSys}, pages 152--160. {ACM}, 2017.

\bibitem{AFM}
Jun Xiao, Hao Ye, Xiangnan He, Hanwang Zhang, Fei Wu, and Tat{-}Seng Chua.
\newblock Attentional factorization machines: Learning the weight of feature
  interactions via attention networks.
\newblock In Carles Sierra, editor, {\em {IJCAI}}, pages 3119--3125. ijcai.org,
  2017.

\bibitem{hjmgraph}
Junmei Hao, Yujie Dun, Guoshuai Zhao, Yuxia Wu, and Xueming Qian.
\newblock Annular-graph attention model for personalized sequential
  recommendation.
\newblock {\em IEEE Transactions on Multimedia}, pages 1--1, 2021.

\bibitem{NGCF}
Xiang Wang, Xiangnan He, Meng Wang, Fuli Feng, and Tat-Seng Chua.
\newblock Neural graph collaborative filtering.
\newblock In {\em SIGIR}, pages 165--174, 2019.

\bibitem{LRGCCF}
Lei Chen, Le~Wu, Richang Hong, Kun Zhang, and Meng Wang.
\newblock Revisiting graph based collaborative filtering: {A} linear residual
  graph convolutional network approach.
\newblock In {\em {AAAI}}, pages 27--34. {AAAI} Press, 2020.

\bibitem{LightGCN}
Xiangnan He, Kuan Deng, Xiang Wang, Yan Li, Yong{-}Dong Zhang, and Meng Wang.
\newblock Lightgcn: Simplifying and powering graph convolution network for
  recommendation.
\newblock In {\em {SIGIR}}, pages 639--648. {ACM}, 2020.

\bibitem{DGCF}
Xiang Wang, Hongye Jin, An~Zhang, Xiangnan He, Tong Xu, and Tat-Seng Chua.
\newblock Disentangled graph collaborative filtering.
\newblock In {\em SIGIR}, pages 1001--1010, 2020.

\bibitem{BPR}
Steffen Rendle, Christoph Freudenthaler, Zeno Gantner, and Lars
  Schmidt{-}Thieme.
\newblock {BPR:} bayesian personalized ranking from implicit feedback.
\newblock In {\em {UAI}}, pages 452--461, 2009.

\bibitem{simCLR}
Ting Chen, Simon Kornblith, Mohammad Norouzi, and Geoffrey~E. Hinton.
\newblock A simple framework for contrastive learning of visual
  representations.
\newblock In {\em {ICML}}, volume 119, pages 1597--1607. {PMLR}, 2020.

\bibitem{moco}
Kaiming He, Haoqi Fan, Yuxin Wu, Saining Xie, and Ross~B. Girshick.
\newblock Momentum contrast for unsupervised visual representation learning.
\newblock In {\em {CVPR}}, pages 9726--9735. {IEEE}, 2020.

\bibitem{supcl}
Prannay Khosla, Piotr Teterwak, Chen Wang, Aaron Sarna, Yonglong Tian, Phillip
  Isola, Aaron Maschinot, Ce~Liu, and Dilip Krishnan.
\newblock Supervised contrastive learning.
\newblock In {\em NeurIPS}, 2020.

\bibitem{pcl}
Junnan Li, Pan Zhou, Caiming Xiong, and Steven C.~H. Hoi.
\newblock Prototypical contrastive learning of unsupervised representations.
\newblock In {\em {ICLR}}. OpenReview.net, 2021.

\bibitem{BYOL}
Jean{-}Bastien Grill, Florian Strub, Florent Altch{\'{e}}, Corentin Tallec,
  Pierre~H. Richemond, Elena Buchatskaya, Carl Doersch, Bernardo~{\'{A}}vila
  Pires, Zhaohan Guo, Mohammad~Gheshlaghi Azar, Bilal Piot, Koray Kavukcuoglu,
  R{\'{e}}mi Munos, and Michal Valko.
\newblock Bootstrap your own latent - {A} new approach to self-supervised
  learning.
\newblock In {\em NeurIPS}, 2020.

\bibitem{GraphCL}
Yuning You, Tianlong Chen, Yongduo Sui, Ting Chen, Zhangyang Wang, and Yang
  Shen.
\newblock Graph contrastive learning with augmentations.
\newblock In {\em NeurIPS}, 2020.

\bibitem{chopra2005learning}
Sumit Chopra, Raia Hadsell, and Yann LeCun.
\newblock Learning a similarity metric discriminatively, with application to
  face verification.
\newblock In {\em CVPR}, volume~1, pages 539--546. IEEE, 2005.

\bibitem{DBLP:conf/cvpr/HadsellCL06}
Raia Hadsell, Sumit Chopra, and Yann LeCun.
\newblock Dimensionality reduction by learning an invariant mapping.
\newblock In {\em {CVPR}}, pages 1735--1742. {IEEE} Computer Society, 2006.

\bibitem{cltmm}
Xulin Song and Zhong Jin.
\newblock Robust label rectifying with consistent contrastive-learning for
  domain adaptive person re-identification.
\newblock {\em IEEE Transactions on Multimedia}, pages 1--1, 2021.

\bibitem{SGL}
Jiancan Wu, Xiang Wang, Fuli Feng, Xiangnan He, Liang Chen, Jianxun Lian, and
  Xing Xie.
\newblock Self-supervised graph learning for recommendation.
\newblock In {\em {SIGIR}}, pages 726--735. {ACM}, 2021.

\bibitem{ContrastiveSequential}
Xu~Xie, Fei Sun, Zhaoyang Liu, Jinyang Gao, Bolin Ding, and Bin Cui.
\newblock Contrastive pre-training for sequential recommendation.
\newblock {\em CoRR}, abs/2010.14395, 2020.

\bibitem{schroff2015facenet}
Florian Schroff, Dmitry Kalenichenko, and James Philbin.
\newblock Facenet: A unified embedding for face recognition and clustering.
\newblock In {\em CVPR}, pages 815--823, 2015.

\bibitem{DBLP:journals/jmlr/ChechikSSB10}
Gal Chechik, Varun Sharma, Uri Shalit, and Samy Bengio.
\newblock Large scale online learning of image similarity through ranking.
\newblock {\em J. Mach. Learn. Res.}, 11:1109--1135, 2010.

\bibitem{weinberger2006distance}
Kilian~Q. Weinberger, John Blitzer, and Lawrence~K. Saul.
\newblock Distance metric learning for large margin nearest neighbor
  classification.
\newblock In {\em NeurIPS}, pages 1473--1480, 2005.

\bibitem{sohn2016improved}
Kihyuk Sohn.
\newblock Improved deep metric learning with multi-class n-pair loss objective.
\newblock In {\em NeurIPS}, pages 1857--1865, 2016.

\bibitem{oord2018representation}
A{\"{a}}ron van~den Oord, Yazhe Li, and Oriol Vinyals.
\newblock Representation learning with contrastive predictive coding.
\newblock {\em CoRR}, abs/1807.03748, 2018.

\bibitem{hjelm2018learning}
R.~Devon Hjelm, Alex Fedorov, Samuel Lavoie{-}Marchildon, Karan Grewal, Philip
  Bachman, Adam Trischler, and Yoshua Bengio.
\newblock Learning deep representations by mutual information estimation and
  maximization.
\newblock In {\em {ICLR}}. OpenReview.net, 2019.

\bibitem{bachman2019learning}
Philip Bachman, R.~Devon Hjelm, and William Buchwalter.
\newblock Learning representations by maximizing mutual information across
  views.
\newblock In {\em NeurIPS}, pages 15509--15519, 2019.

\bibitem{Nonpara}
Zhirong Wu, Yuanjun Xiong, Stella~X. Yu, and Dahua Lin.
\newblock Unsupervised feature learning via non-parametric instance
  discrimination.
\newblock In {\em {CVPR}}, pages 3733--3742. {IEEE} Computer Society, 2018.

\bibitem{chen2017sampling}
Ting Chen, Yizhou Sun, Yue Shi, and Liangjie Hong.
\newblock On sampling strategies for neural network-based collaborative
  filtering.
\newblock In {\em KDD}, pages 767--776, 2017.

\bibitem{GCC}
Jiezhong Qiu, Qibin Chen, Yuxiao Dong, Jing Zhang, Hongxia Yang, Ming Ding,
  Kuansan Wang, and Jie Tang.
\newblock {GCC:} graph contrastive coding for graph neural network
  pre-training.
\newblock In {\em {KDD}}, pages 1150--1160. {ACM}, 2020.

\bibitem{MVGCL}
Kaveh Hassani and Amir Hosein~Khas Ahmadi.
\newblock Contrastive multi-view representation learning on graphs.
\newblock In {\em {ICML}}, volume 119, pages 4116--4126. {PMLR}, 2020.

\bibitem{GAA}
Yanqiao Zhu, Yichen Xu, Feng Yu, Qiang Liu, Shu Wu, and Liang Wang.
\newblock Graph contrastive learning with adaptive augmentation.
\newblock In {\em {WWW}}, pages 2069--2080. {ACM} / {IW3C2}, 2021.

\bibitem{Gaug}
Tong Zhao, Yozen Liu, Leonardo Neves, Oliver~J. Woodford, Meng Jiang, and Neil
  Shah.
\newblock Data augmentation for graph neural networks.
\newblock In {\em {IAAI}}, pages 11015--11023. {AAAI} Press, 2021.

\bibitem{softcl}
Janine Thoma, Danda~Pani Paudel, and Luc~Van Gool.
\newblock Soft contrastive learning for visual localization.
\newblock In {\em NeurIPS}, 2020.

\bibitem{DebiasedCL}
Ching{-}Yao Chuang, Joshua Robinson, Yen{-}Chen Lin, Antonio Torralba, and
  Stefanie Jegelka.
\newblock Debiased contrastive learning.
\newblock In {\em NeurIPS}, 2020.

\bibitem{GCMC}
Rianne van~den Berg, Thomas~N. Kipf, and Max Welling.
\newblock Graph convolutional matrix completion.
\newblock {\em CoRR}, abs/1706.02263, 2017.

\bibitem{PinSage}
Rex Ying, Ruining He, Kaifeng Chen, Pong Eksombatchai, William~L. Hamilton, and
  Jure Leskovec.
\newblock Graph convolutional neural networks for web-scale recommender
  systems.
\newblock In {\em {KDD}}, pages 974--983, 2018.

\bibitem{multiVAE}
Dawen Liang, Rahul~G. Krishnan, Matthew~D. Hoffman, and Tony Jebara.
\newblock Variational autoencoders for collaborative filtering.
\newblock In {\em {WWW}}, pages 689--698. {ACM}, 2018.

\bibitem{graphmlp}
Yang Hu, Haoxuan You, Zhecan Wang, Zhicheng Wang, Erjin Zhou, and Yue Gao.
\newblock Graph-mlp: Node classification without message passing in graph.
\newblock {\em CoRR}, abs/2106.04051, 2021.

\end{thebibliography}
  
\end{document}